# Size-dependent multiexciton dynamics governs scintillation from perovskite quantum dots

Andrea Fratelli[1], Matteo L. Zaffalon[1], Emanuele Mazzola[2], Dmitry Dirin[3], Ihor Cherniukh[3], Clara Otero Martínez[4], Matteo Salomoni[2], Francesco Carulli[1], Francesco Meinardi[1], Luca Gironi[2,5], Liberato Manna[4*], Maksym V. Kovalenko[3*], Sergio Brovelli[1,5*]

[1] *Dipartimento di Scienza dei Materiali, Università degli Studi di Milano-Bicocca, Via R. Cozzi 55, 20125, Milano, Italy*

[2] *Dipartimento di Fisica, Università degli Studi di Milano-Bicocca, Piazza della Scienza 3, 20126, Milano, Italy*

[3] *Department of Chemistry and Applied Bioscience, ETH Zürich, Zürich, Switzerland.*
*Laboratory for Thin Films and Photovoltaics and Laboratory for Transport at Nanoscale Interfaces, Empa – Swiss Federal Laboratories for Materials Science and Technology, Dübendorf, Switzerland.*

[4] *Nanochemistry, Istituto Italiano di Tecnologia, Via Morego 30, Genova, Italy*

[5] *INFN - Sezione di Milano - Bicocca, Milano I-20126 – Italy*

**The recent emergence of quantum confined nanomaterials in the field of radiation detection, in particular lead halide perovskite nanocrystals, offers potentially revolutionary scalability and performance advantages over conventional materials. This development raises fundamental questions about the mechanism of scintillation itself at the nanoscale and the role of particle size, arguably the most defining parameter of quantum dots. Understanding this is crucial for the design and optimisation of future nanotechnology scintillators. In this work, we address these open questions by theoretically and experimentally studying the size-dependent scintillation of $CsPbBr_3$ nanocrystals using a combination of Monte Carlo simulations, spectroscopic, and radiometric techniques. The results reveal and unravel a complex parametric space where the fine balance between the simultaneous effects of size-dependent energy deposition, (multi-)exciton population, and light emission under ionizing excitation, typical of confined particles, combine to maximize the scintillation efficiency and time performance of larger nanocrystals due to greater stopping power and reduced Auger decay. The remarkable agreement between theory and experiment produces a fully validated descriptive model that unprecedentedly predicts the scintillation yield and kinetics of nanocrystals without free parameters, providing the first fundamental guide for the rational design of nanoscale scintillators.**

Ionising radiation detection is critical across diverse domains such as precision medicine[1, 2], industrial[3] and national security[4], environmental monitoring[5], energy management[5], and cutting-edge scientific research in high-energy physics[6] (HEP) at particle accelerators in the search for rare events in nuclear physics. In these applications, scintillator materials play a crucial role by converting ionising radiation into detectable signals using high-performance photosensors such as phototubes or silicon photomultipliers (SiPMs). Ideal scintillators should feature compositions rich in high atomic number ($Z$) elements, ensuring a high probability of interaction with radiation (proportional to $Z^n$, $n$=1-5 depending on the radiation type and interaction process)[7, 8], along with high density, stability to radiation (so-called radiation hardness), efficiency, and speed of emission processes, which is especially vital in time-of-flight-based technologies like positron emission tomography[9] (ToF-PET) and high-brightness beam detection[10, 11]. The figure of merit for ToF-PET is the coincidence time resolution[12], $CTR = 3.33\sqrt{(\tau_{RISE} \times \tau_{EFF})/\Gamma}$, where $\tau_{RISE}$ is the signal risetime (typically due to the detection chain) and $\tau_{EFF} = \left(\sum_i \frac{R_i}{\tau_i}\right)^{-1}$ is the effective scintillation lifetime[11] (rate $k_{EFF}=\frac{1}{\tau_{EFF}}$) obtained, in case of multi-exponential decay kinetics, as the harmonic average of the *i*-th scintillation decay components weighted by their respective time-integrated relative contributions $R_i$. The term $\Gamma = \Phi_{SCINT} \times E \times \beta \times \chi$ is the intensity of the scintillation signal that depends on the scintillation efficiency ($\Phi_{SCINT}$), on the amount of energy ($E$) deposited in the scintillator, on the light outcoupling efficiency ($\beta$), and on the quantum efficiency of the coupled photodetector ($\chi$). The quantity $LY = \Phi_{SCINT} \times E \times \beta$ is commonly referred to as the light yield and corresponds to the number of emitted photons per unit energy deposited[7]. In the ToF-PET field, the main challenge is to achieve CTR≤10 ps that would significantly reduce the acquisition time (most accurate commercial devices feature CTR values around 200 ps) while improving the signal-to-noise ratio[2, 10], leading to millimetre spatial resolution in cancer diagnostics and providing high image quality at reduced doses. In the case of HEP experiments, the push to explore the limits of the Standard Model at the frontiers of energy and intensity requires experiments to operate at ever-higher rates[5]. Scintillation detectors for the High-Luminosity Large Hadron Collider (HL-LHC) and Future Circular Collider (FCC) era will require time resolutions on the order of a few tens of ps or less, with double-pulse separation at the level of a few ns. In this case, there is no exhaustive figure of merit regarding the timing performance as with CTR but, similarly to ToF-PET, the timing resolution is given by the variance with which enough photoelectrons are collected to provide a statistically relevant signal (hence often quantified in terms of photons MeV$^{-1}$ ns$^{-1}$). In both fields, the main obstacle to achieving the required energy and temporal resolution lies in the limitations of scintillator materials that are often chosen based on a trade-off between their performance, cost, and availability. Inorganic crystals[13] offer high efficiency and energy resolution but are slow, expensive, and difficult to mass-produce, whereas plastic scintillators[14] are cost-effective and fast emitting but suffer from low density, efficiency and radiation hardness. To address the drawbacks of both types and capitalize on their strengths, nanocomposite scintillators have emerged recently[15, 16]. These scintillators feature optical-grade plastic matrices as the waveguiding component, while high-Z nanocrystal (NCs) synthesized using scalable chemical techniques provide scintillation[15, 17, 18]. Importantly, using NCs as nanoscintillators in host matrices not only offers a solution to overcome the scalability limitations of conventional materials but also avenue to enhance the scintillation performance. This is due to the unique photophysics of quantum-confined materials, providing size-tunable emission spectra that match perfectly with the spectral sensitivity of light detectors and ultrafast sub-nanosecond scintillation kinetics resulting from recombination of multi-exciton generated upon interaction with ionising radiation, as demonstrated recently across various classes of NCs[16, 18-20].

One category of nanomaterials that has garnered particular attention within this context is lead halide perovskite NCs (LHP-NCs)[21, 22, 23, 24], with $CsPbBr_3$ emerging as the dominant player[21, 25, 26]. These materials feature a composition based on heavy metals, remarkable resistance to radiation[18, 23, 27], extensive scalability facilitated by low-temperature methods[28, 29], and efficient, fast scintillation[19, 24, 30] owing to the unique tolerance to structural defects[31]. In recent years, there has been a growing body of research aimed at optimizing their scintillation and developing nanocomposites for various radiation detection applications[32], particularly X-ray detectors and screens utilized in medical imaging and object inspection. Time-resolved scintillation studies have further demonstrated ultrafast radioluminescence (RL) kinetics due to substantial contributions by the recombination of biexcitons (indicated as $XX$; $X$ denotes single exciton species) generated upon interaction with ionising radiation, which is particularly promising for fast-timing applications[18, 19]. To fully exploit the potential of nanoscale materials for radiation detection, it is essential to fully understand and control the key parameters that govern the scintillation processes, which, as we demonstrate below, are strongly size-dependent. For example, the particle size has a strong influence on the amount of energy deposited within a single NC after ionising excitation, resulting in size-dependent exciton occupancy. The resulting multiexciton scintillation is also affected by nonradiative Auger recombination (AR) – that is the nonradiative annihilation of one exciton in favour of a third carrier[33] – whose rate increases with the inverse of the particle volume ($k_{AR} \propto V^{-1}$)[34]. Overall, this leads to a complex interplay between size effects on the scintillation yield and time kinetics that requires a detailed understanding for proper material optimization, which has not been investigated to date.

Here we aim to fill this gap by studying the effect of particle size on the scintillation efficiency and kinetics of $CsPbBr_3$ NCs ranging from $d$=3 nm (lateral size) to $d$=15 nm, with tunable emission from 470 nm to 520 nm. The dependence of scintillation intensity of single NCs ($I_{SCINT}^{NC}$) and NC ensembles ($I_{SCINT}^{ENS}$) and $\tau_{EFF}$ on particle size is first theoretically analysed through the combination of the emission rate equations in the multiexciton regime and Geant4 Monte Carlo simulations[35], and then experimentally validated, yielding an intertwined parameter space where the size-dependent initial exciton population per NC (denoted as $\langle N \rangle$), the AR rate and the fluorescence efficiency are the key elements. We have experimentally evaluated all the key parameters necessary to describe the scintillation mechanisms using a combination of optical spectroscopies and radiometric experiments as summarized in **Scheme 1a**. X-ray attenuation and scintillation experiments, confirmed by Monte Carlo simulations, showed that the stopping power of NC ensembles with the same total mass is independent of particle size, while the energy deposition within a single NC increases with size, closely resembling the carrier multiplication phenomenology. This leads to substantially higher $\langle N \rangle$-values for larger particles making high-order exciton contributions gradually more relevant, as confirmed by time resolved radioluminescence (RL) experiments. Consistent with the literature[36-38], AR was found to be efficient in all NC samples, resulting in scintillation dominated by single-exciton photoluminescence (PL) efficiency ($\Phi_X$) and progressive acceleration of $\tau_{EFF}$ with increasing NC size due to reduced AR quenching of the ultrafast $XX$ decay. The whole body of experimental data validates, with no free parameters, the theoretical model, which disentangles the single-particle and ensemble effects on the size dependence of NCs scintillation and provides unprecedented guidelines for tailored technological optimization of nanoscale materials in radiation detection.

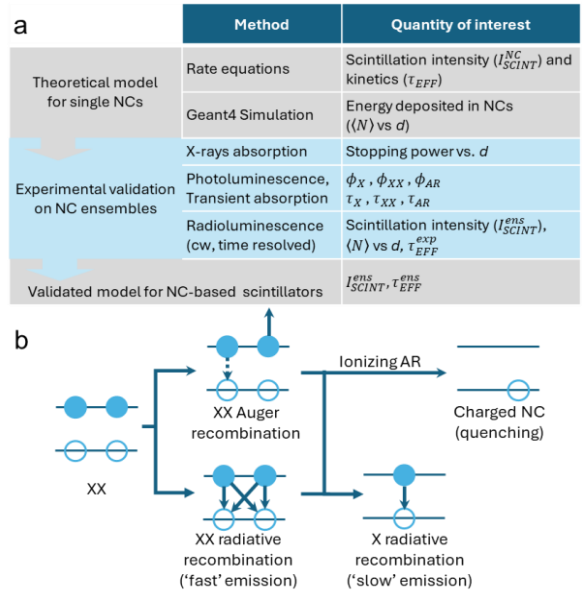

**Scheme 1. a.** Theoretical and experimental approaches for a validated model for the scintillation of NCs. **b**. Schematic depiction of the possible decay channels for biexcitons created upon ionising excitation.

**Results and discussion**
To accurately describe size effects on NC-based scintillators, it is necessary to consider both single-particle and ensemble effects. Single-particle effects concern the response of individual NCs to ionising radiation, such as interaction property, stopping power, and AR-dependent emission efficiency. Ensemble effects, on the other hand, are more technological and are crucial for the engineering of NCs-based scintillators. Consider, for example, that for the same mass of scintillator material ($M$), a nanocomposite scintillator contains an amount of NCs ($n_{NC}$) that is proportional to the inverse of their volume ($n_{NC} = M/\rho V_{NC}$, where $\rho$ is the density), which can introduce a large scaling factor for single-particle characteristics (e.g. $n_{NC}(d = 3\ nm)/n_{NC}(d = 10\ nm) \sim 30$). We therefore begin with the single-NC treatment, which will then be implemented with ensemble considerations to provide us with a realistic interpretative model of the experimental findings. As the purpose of this work is to provide guidelines for NC selection and design, we focus on isolated non-interacting particles and leave the extension of our treatment to dense NC solids to a dedicated study.

*Theoretical considerations on the impact of the NC size on the scintillation parameters.* A discussion of the role of size in the scintillation of NCs requires addressing fundamental aspects of the NC photophysics under ionising excitation that determines high-order exciton populations subject to AR. For the sake of this discussion, we do not consider nonradiative decay pathways other than AR (e.g. trapping, multi-phonon relaxation, that is, we consider a X emission efficiency of $\Phi_X =1$) and neglect the case of trions as they have been experimentally shown to give a relatively minor contribution to the scintillation kinetics[19] and whose formation is situational as it largely depends on trapping processes. In the absence of AR, *XX* radiatively decay into a *X* species via the emission of a photon with an accelerated radiative rate due to increased emission statistics ($k_{XX}=4k_X$ where $k_X$ is the *X* radiative rate[33]), giving rise to a bi-exponential scintillation decay profile consisting of a fast *XX* component followed by a slower one due to the resulting *X* (see **Scheme 1b**)[18, 19, 29, 39]. Auger decay involves the annihilation of a first exciton and the simultaneous promotion of a second carrier (belonging to the second exciton) to an energy equal to twice the energy gap. If this additional energy exceeds the ionisation energy of the material, AR ionises the NC and quenches the emission completely[40]. Alternatively, in the non-ionising AR case, the hot carrier rapidly thermalizes back to the band edge, constituting a secondary *X*. In the absence of other nonradiative processes, such as trapping of hot carriers, as it occurs in B-type blinking[41], this results in a net loss of half the photons that would be emitted if both excitons recombined radiatively. This is relevant for addressing the scintillation kinetics because the acceleration of the *XX* contribution is accompanied by the repopulation of *X* species that modify $\tau_{EFF}$ and partially compensate the light loss. The interplay between these recombination mechanisms of individual NCs in the *XX* regime is better understood considering the characteristic rate equations under instantaneous excitation (we hereby describe all processes through their rate for clarity),

$$\dot{N}_X = -N_X^0 k_X + N_{XX}^0(k_{XX} + k_{AR}\xi) \quad (1)$$
$$\dot{N}_{XX} = -N_{XX}^0(k_{XX} + k_{AR}) \quad (2)$$

Where $N_X^0$ and $N_{XX}^0$ are the initial *X* and *XX* Poissonian populations following primary ionising excitation obtained as:

$$N_X^0 = \langle N \rangle^2 e^{-\langle N \rangle}$$
$$N_{XX}^0 = \langle N \rangle \times (1 - \langle N \rangle e^{-\langle N \rangle} - e^{-\langle N \rangle})$$

The population of *X* generated indirectly by the decay of *XX* via a radiative pathway or following AR is accounted for by the positive term in **Eq.1**. The probability of undergoing ionisation following AR is expressed by the term $(1 - \xi)$. The solutions to **Eq.1** and **Eq.2** are, respectively,

$$N_X(t) = \left[N_X^0 + N_{XX}^0 \frac{k_{AR}\xi + k_{XX}}{k_{XX}+k_{AR}-k_X}\left(1 - e^{-(k_{XX}+k_{AR}-k_X)t}\right)\right]e^{-(k_X)t} \quad (3)$$

$$N_{XX}(t) = N_{XX}^0 e^{-(k_{XX}+k_{AR})t} \quad (4)$$

Solving **Eq.1** and **Eq.2** in the steady state gives the analytic expression of $I_{SCINT}^{NC}$ emitted by a single NC particle following the creation of $\langle N \rangle$ excitons by an ionising photon:

$$I_{SCINT}^{NC}(\langle N \rangle) \propto \Phi_X(N_X^0 + N_{XX}^0\Phi_{XX} + N_{XX}^0\Phi_{AR}\xi) + N_{XX}^0\Phi_{XX}, \quad (5)$$

where the *X* and *XX* processes contribute through their respective $\Phi_X$, $\Phi_{XX} = 4k_X/(4k_X + k_{AR})$ and relative populations, which in turn depend on $\Phi_{AR} = k_{AR}/(4k_X + k_{AR})$. Essentially, **Eq.5** allows for disentangling the *X* and *XX* contributions to the scintillation of single NCs which depend on the NC size through the corresponding $\Phi_{AR}$ and $\langle N \rangle$. Similarly, **Eq.3** and **Eq.4** describe the time kinetics of the scintillation process and allow the determination of $\tau_{EFF}$ in a similar way to the experimental data. Together with **Eq.5**, this enables predicting the single particle timing performance through the quantity, $CTR_{NC} \propto \sqrt{\frac{\tau_{EFF}}{I_{SCINT}}}$. We underline that **Eq.1-5** are valid for any excitation source and thus hold also for any optoelectronic/photonic application involving the recombination of biexcitons.

The simulated $\tau_{EFF}$, $I_{SCINT}^{NC}$, and $CTR_{NC}$ values computed using $k_X = 10^{-4}\ ps^{-1}$ (corresponding to a *X* lifetime of 10 ns typical of most UV-Vis emitting NCs), $\Phi_{AR} = 0 - 0.99$ and $\xi = 0.5$ are shown in **Figure 1a** (see Supporting Figure 3 for the simulations with $\xi = 0$ and 1), where we have treated all excitons with order higher than one as *XX*. Increasing $\langle N \rangle$ from 0.3 to 3.3, which is associated to situations in which larger amounts of energy are deposited within a NC, appears to be the dominant effect and results in the gradual enhancement of $I_{SCINT}^{NC}$. More efficient AR reduces the effect by quenching the ever more dominant higher-order excitons. The evolution of $\tau_{EFF}$ is more complex. On the one hand, it gradually accelerates with increasing $\langle N \rangle$ due to the increasing $XX$ contributions, and on the other hand, it follows a non-monotonic trend with the AR efficiency, which initially accelerates $\tau_{EFF}$ by speeding up the $XX$ decay, but as it approaches unity completely suppresses the $XX$ contribution, resulting in a much slower kinetics, essentially determined solely by the $X$ decay. As a result, $CTR_{NC}$ improves (accelerates) substantially with $\langle N \rangle$ and is only weakly worsened by AR. In fact, the crucial aspect that emerges from this analysis is the importance of $\langle N \rangle$ generated as a result of ionising interaction in both light output and timing, pointing to the technological relevance of maximizing the single particle energy retention, while NC engineering for suppressing AR is relevant for maximizing the scintillation intensity but plays a relatively minor role in the timing performance.

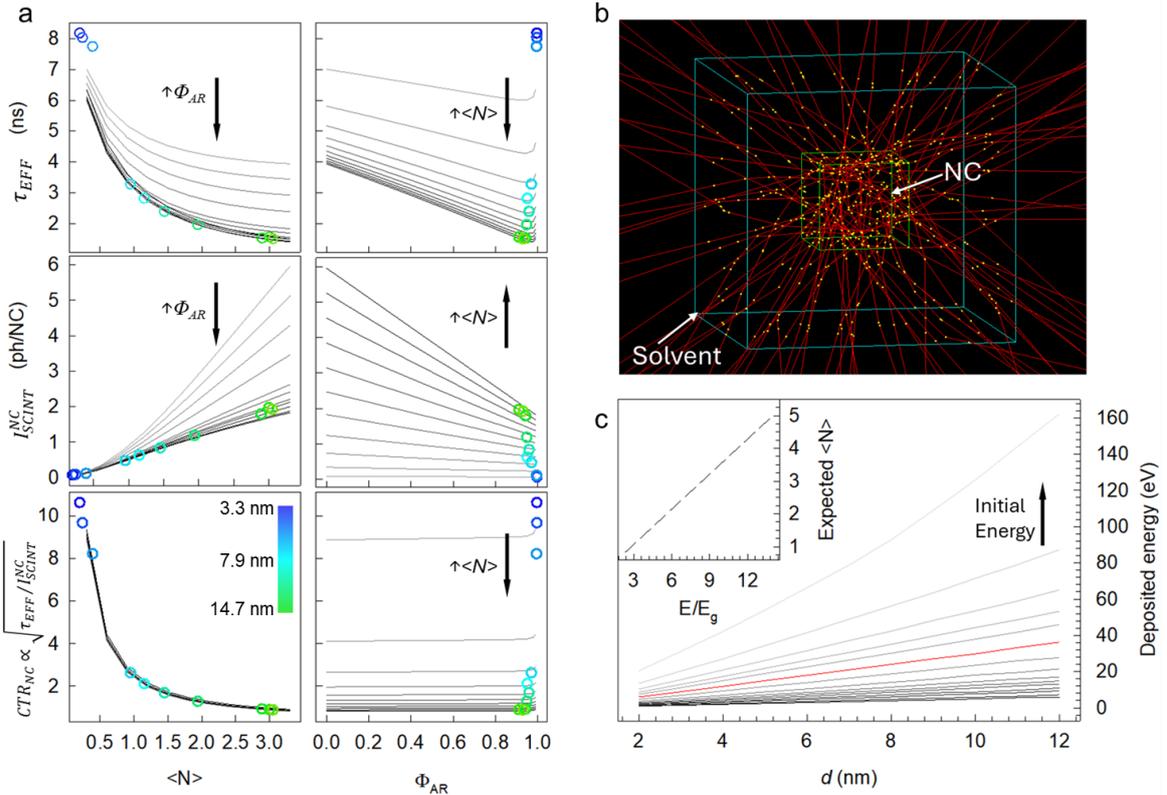

**Figure 1. a.** Simulation of the effective lifetime ($\tau_{EFF}$), the single-particle scintillation intensity ($I_{SCINT}^{NC}$), and the square root of their ratio ($\propto CTR_{NC}$) as a function of the particle size for increasing values of $\langle N \rangle$ (0.3-3.3) or $\Phi_{AR}$ (0-0.99). Arrows indicate increasing $\Phi_{AR}$ or $\langle N \rangle$. Coloured circles represent the experimental values obtained for CsPbBr$_3$ NCs with increasing size from 3.3 nm to 14.7 nm (from blue to green, legend in the bottom panel) **b.** Schematic representation of the simulation using Geant4 of the energy release inside a single NC following a photoelectric event with E=7 keV. **c.** Results of the simulation of deposited energy with respect to the NC size, the red line corresponds to E=7 keV. The black arrow indicates increasing initial electron energy (E=1-100 keV). Inset: simulation of the expected $\langle N \rangle$ with respect to the deposited energy (expressed as number of bandgaps) computed considering the bulk electron-hole pair formation energy $\varepsilon_{e\text{-}h} \sim 2.8 E_g$.[42]

Therefore, to investigate the role of size on the energy retention capability of NCs (responsible for $\langle N \rangle$), we computed the fraction of energy deposited within single NCs of increasing size (from $d$=2 to 15 nm) following a photoelectric event producing electrons with energy $E$=1-100 keV (schematic representation of a representative case with E=7 keV is shown in **Figure 1b**). Interestingly, the Geant4 simulation[35] shows that for any initial energy, in the investigated size range, the energy deposited within a single NC grows with $d$ (**Figure 1c**), with the smallest NCs releasing almost all of the initial electron energy into the outer medium. In the case of $E = 7$ keV (corresponding to the average energy of the X-rays used in our experiments), the energy deposited ranges from 5 eV to 35 eV for the smallest and the largest CsPbBr$_3$ NCs, corresponding to 2-13 times the corresponding energy gap ($E_g$). This trend highlights the instructive analogy between the excitation phase of the scintillation process (which gives rise to ⟨$N$⟩) and carrier multiplication[43], a process that has been studied in detail in NCs[44], where the number of band edge excitons produced scales linearly with the absorbed energy divided by the absorber $E_g$, and the angular coefficient is the inverse of the electron-hole pair formation energy ($\varepsilon_{e-h}$, in bulk semiconductors $\varepsilon_{e-h} \sim 2.8 E_g$)[42]. The inset of **Figure 1c** shows the expected trend for an example case where $\varepsilon_{e-h}$ is fixed at the bulk limit, motivated by the fact that the NCs states populated under high energy radiation are not subject to quantum confinement. As demonstrated below, this predicted ⟨$N$⟩-trend agrees remarkably well with the experimental results, enabling us to match the experimental $\tau_{EFF}$, $I_{SCINT}^{NC}$, and $CTR_{NC}$-values with the respective curves in **Figure 1a** (highlighted as coloured dots).

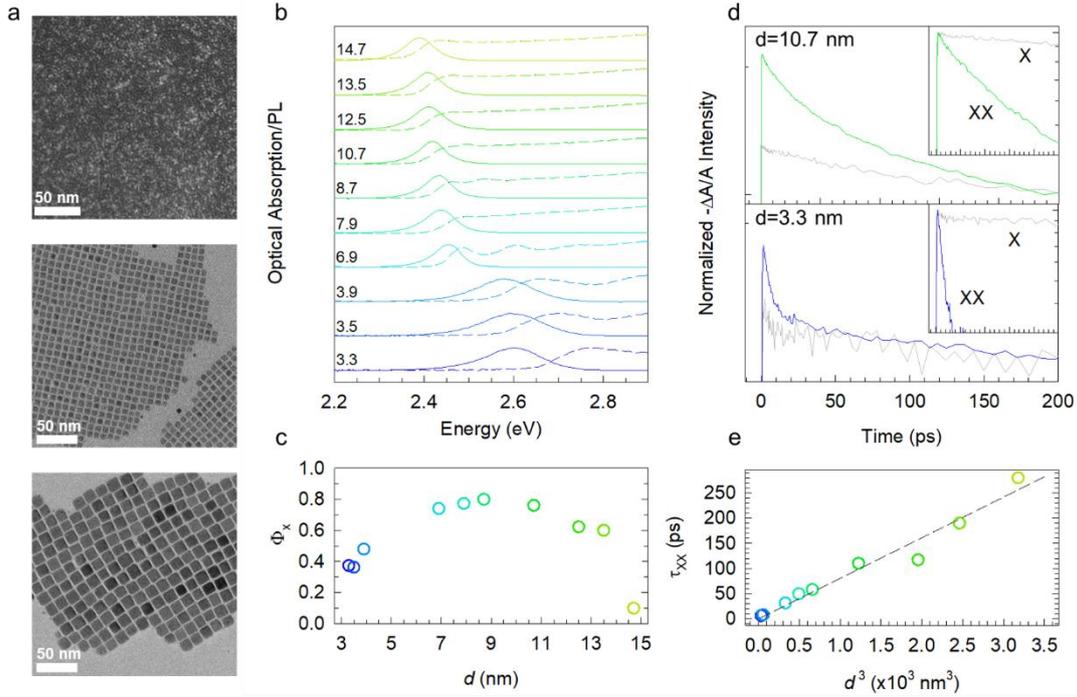

**Figure 2. a.** HRTEM images for CsPbBr$_3$ NCs with $d$=3.9 nm, $d$=6.9 nm, $d$=13.5 nm. **b.** Optical absorption and PL spectra of the complete sample set (respective NC size in nm is indicated). **c.** PL quantum efficiency values estimated in single exciton regime with respect to the NC size. **d.** TA dynamics for 10.7 nm and 3.3 nm NC for ⟨$N$⟩ = 0.16 (grey lines) and ⟨$N$⟩ = 1.8 (green, blue line) at their respective 1S bleach maximum. Inset: single exciton and biexciton components (time axis 0-100 ps with major ticks every 20 ps) extracted by the successive subtraction method[33]. **e.** Biexciton lifetimes, $\tau_{XX}$, as a function of the NC volume. Dashed line, linear interpolation highlighting the volume scaling.

*Spectroscopic and scintillation experiments on CsPbBr$_3$ NCs of increasing size.* To experimentally validate the theoretical framework just presented we performed optical spectroscopy and RL experiments on CsPbBr$_3$ NCs of increasing size from ~3 nm to ~15 nm. The particles were of cubic shape and orthorhombic crystalline phase, as shown in the high-resolution transmission electron microscope images of representative samples in **Figure 2a** and respective X-ray diffraction patterns (Supporting Figure 1,2). The corresponding optical absorption and PL spectra are reported in **Figure 2b**, showing the progressive blue-shift for decreasing NC sizes due to increasing quantum confinement[45]. The PL yield (measured at low excitation fluence to ensure a $X$ photophysics, **Figure 2c**) spanned non-monotonically from $\Phi_X$=12% to 80%, with the smallest and largest particles systematically showing lower efficiency, consistent with common observation of better surface passivation for intermediate sized NCs (i.e. $7\ nm \lesssim d \lesssim 11$ nm). The respective PL dynamics (Supporting Figure 4) showed single exciton effective lifetimes around 10 ns in all cases, with a measurable lengthening for the largest NCs consistent with the stronger symmetry forbidden s-p character of the radiative transition in large particles[46]. We next interrogated the photophysics *vs.* size in the multi-exciton regime by performing TA measurements as a function of increasing excitation fluence[36-38]. The 1S bleach dynamics are shown in **Figure 2d** for representative NC samples with $d$=10.7 nm and 3.3 nm (the whole set of data is reported in Supporting Figure 6). At low excitation fluence (i.e. the $X$ regime), all samples showed the characteristic peak at their respective band-edge energy due to bleaching of the 1S exciton absorption at all times[36] (Supporting Figure 5). The corresponding time dynamics was essentially single exponential with characteristic time, $\tau_X$~8 ns, matching the corresponding PL kinetics. Upon increasing the excitation fluence, a low energy shoulder appeared in the TA spectra, consistent with the attractive character of $XX$ in CsPbBr$_3$ NCs[36, 38] and the TA dynamics developed an initial ultrafast component due to $XX$ decay with amplitude following the Poisson biexciton state-filling statistics[47] (~$\langle N \rangle^2$, Supporting Figure 7). Higher excitation fluences gave rise to a high energy shoulder with even more accelerated decay due to higher order multiexcitons. To extract the $XX$ and AR rates ($k_{XX}, k_{AR}$) and corresponding efficiencies ($\Phi_{XX}, \Phi_{AR}$), we normalized the TA dynamics to their slow $X$ tail and progressively subtracted curves with increasing $\langle N \rangle$[36, 38]. The obtained $XX$ decay curves are shown in the inset of **Figure 2d** (and in Supporting Figure 8). The $XX$ lifetimes of the whole sample set are shown in **Figure 2e** following the universal volume scaling law[36, 37]. The corresponding $\Phi_{AR}$-values spanned from ~90% for the largest particles to ~99% for strongly confined NCs, in good agreement with the literature[36, 37, 48].

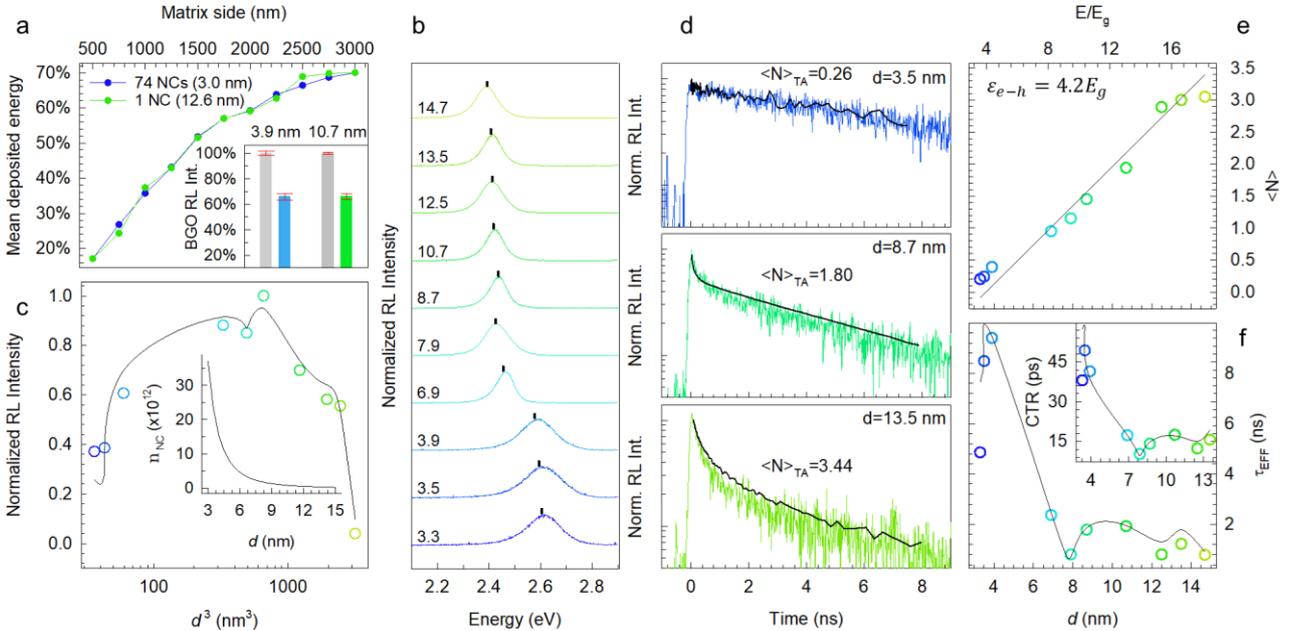

**Figure 3. a**. Geant4 simulation of the energy deposition in two samples containing the mass of CsPbBr$_3$ in the form of one NC with $d$=12.6 nm or 74 NCs with $d$=3 nm (identical total volume of 2000 nm$^3$) of NCs. Inset: relative RL intensity of BGO in X-ray in the absence (grey) or in the presence of two solutions containing same mass of CsPbBr$_3$ NCs with $d$=3.9 nm (blue bar) or $d$= 10.7 nm (green bar) corresponding in both cases to 34±3% attenuation. **b**. RL spectra as a function of $d$ (in nm as indicated). Thicks indicate the corresponding PL maximum. **c.** Relative RL intensity vs. $d^3$. The solid line is the simulated trend. Inset: Number of NCs in 75 μL of octane solution with OD=0.03 vs. NC size. **d**. Time resolved RL decay curves for the NCs with $d$=3.5, 8.7 or 13.5 nm. The blue lines are the TA kinetics at the corresponding $\langle N \rangle$. **e**. Experimental values of $\langle N \rangle$ with respect to the NC size and to the number of bandgaps predicted by Geant4 (top X-axis). The bottom axis is the particle size $d$ as in 'f'. **f**. Effective lifetime extracted from the fit of time-resolved RL decays together with the simulated trend (solid line). Inset: estimated $CTR$ values obtained setting $\tau_{RISE} = 100 ps$ and considering a LY=10kph/MeV for the $d$=8.7 nm NC sample.

We then proceeded to study the scintillation behaviour in order to experimentally measure the actual exciton occupancy $N$ created upon ionising X-ray excitation and the RL intensity/timing vs. $d$. As a first step, we verified that the X-ray attenuation of NC solutions with a given CsPbBr$_3$ mass content is size-independent, as this is closely related to the RL excitation rate. To this end, we measured the RL of a Bi$_4$Ge$_3$O$_{12}$ (BGO) crystal with and without interposed NC solutions ($d$=3.9 vs. 10.7 nm) with identical band-edge absorption between the scintillator and the X-ray source. The ratio between the acquired RL intensities with and without samples partially blocking the X-ray beam estimated the fraction of transmitted X-rays. In parallel, we used Geant4 simulations to calculate the energy deposited in two samples containing identical amounts of $CsPbBr_3$, one in the form of a single NC with $d$=12.6 nm and the other in the form of 74 NCs with $d$=3 (both having total volume of 2000 nm$^3$). The calculations were also performed for different NC concentrations by simulating varying matrix sizes from 0.125 to 27 μm$^3$. As shown in **Figure 3a**, both the calculations and the experiment confirmed that the total stopping power was independent of the NC size, consistent with the fact that the samples had identical average densities. This enables us to quantitatively compare the RL intensity of NC solution with the same mass concentration. For all investigated samples, the RL spectra shown in **Figure 3b** matched the corresponding PL (the PL peak positions are marked with a tick on top of the corresponding RL curve), indicating that light emission was due to pure excitonic recombination with negligible influence from defect states under both excitation conditions. We then proceeded to measure the relative scintillation intensity and decay kinetics of X-ray isoabsorbing diluted solutions of the $CsPbBr_3$ NC – so as to avoid inter-NC processes – under identical excitation and collection conditions. The integrated relative RL efficiencies are reported in **Figure 3c,** following a similar trend to $\Phi_X$, thus highlighting the key role of the luminescence yield in the single exciton regime in the scintillation output. Time-resolved RL measurements complete the picture by providing direct evidence of the strong size dependence of $\langle N \rangle$ and of the *XX* decay rate and relative contribution under X-ray excitation. Specifically, **Figure 3d** shows the RL decay curves of three representative NC samples with $d$=3.5, 8.7, 13.5 nm (the complete set is shown in Supporting Figure 9). The largest NCs showed multi-exponential kinetics with a fast *XX* component followed by the slower *X* decay[26]. In line with previous reports[18, 19], an intermediate component due to charged excitons was found ranging from 0.1%-17% of the total signal. Most importantly, the relative weight of the *XX* contribution decreased gradually with decreasing $d$, resulting in the smallest particles showing only the *X* decay, which is consistent with the calculated small energy retention resulting in negligible *XX* population. Following the typical approach for ultrafast kinetic studies in NCs, the ratio between the amplitudes of the *X* and *XX* components enables estimation of $\langle N \rangle$ through Poissonian statistic[33, 36]. Notably, as shown in **Figure 3e,** increasing the particle size resulted in the linear growth of $\langle N \rangle$, in remarkable agreement with the single particle energy-retention Monte Carlo simulations shown in the inset of **Figure 1c** and with the corresponding TA trace reported as blue lines in **Figure 3d**, further confirming the correct quantification of the exciton occupancy. Expressing the emerging $\langle N \rangle$ as a function of the number of bandgap energies deposited within a single NC yielded an electron-hole pair formation energy of $\varepsilon_{e-h}$=4.2E$_g$, which is slightly above the bulk limit in carrier multiplication processes as expected for quantum confined particles[43]. Also consistent with the TA data in **Figure 2e**, the *XX* decay component of the RL accelerated with decreasing particle size due to gradually stronger AR which also concomitantly led to a decrease of the respective relative weight. Overall, as shown in **Figure 3f**, this led to the acceleration of the experimental $\tau_{EFF}$ – extracted as the harmonic average of the *X*, *XX* and trion contributions - with increasing particle size despite the corresponding slower *XX* decay, thus further highlighting the relevance of the interaction with ionising radiation over AR. It is important to note the opposite size dependence of the kinetics of scintillation, which accelerates with $d$ due to its strongly biexcitonic nature, and that of PL, which instead slows down in large particles due to the increasing hybridisation of the s and p states, which makes optical transitions symmetry forbidden[46].

Crucially, using the experimental values for all scintillation parameters (i.e. $\langle N \rangle$, decay rates and efficiencies) quantified through the optical and radiometric experiments without free parameters, we were able to use **Eq. 3-5** to model the evolution with the NC size of $I_{SCINT}^{ENS}$ of same-mass ensembles of $CsPbBr_3$ NCs as well as their $\tau_{EFF}$. To do so, we accounted for the different mass distributions across the sample set by scaling the single particle $I_{SCINT}^{NC}$ described by **Eq.5** for the respective number of NCs in the ensemble – estimated by dividing the total mass by the 'molecular' weight of a single NC (inset of **Figure 3c**) – and obtained the scintillation intensity, $I_{SCINT}^{ENS} = I_{SCINT}^{NC} \times n_{NC}$ of NC ensembles vs. $d$. The resulting simulated curve is reported as a solid line in **Figure 3d**, showing excellent match with the experimental values. Similarly, running **Eq.3** and **Eq.4** with the experimentally measured lifetimes and ⟨$N$⟩-values accurately describes the dependence of $\tau_{EFF}$ on the particle size (solid line in **Figure 3f**) thus further confirming the validity of the model providing the first description of the scintillation yield and time kinetics of isolated NCs. Finally, to provide a potentially technologically relevant insight into the timing performance of NCs, it is instructive to evaluate the impact of the NC size on the potential time resolution of more realistic NCs ensembles by calculating the *CTR* considering *LY* matching literature reports on CsPbBr$_3$ NCs – e.g. *LY* =10 kph/MeV for the *d*=8.7 nm sample corresponding to a light output of 514 photons for 511 keV gamma excitation. By setting $\tau_{RISE} = 100\ ps$, we estimated the *CTR* values in the inset of **Figure 3f** improving (decreases) gradually with the NC size and reaching potentially relevant values as fast as 15 ps or so, a trend again correctly predicted by the simulations in **Figure 1a**.

In conclusion, based on the observed trends, the single particle energy retention appears to dominate the scintillation mechanism of NCs, as it determines the initial exciton occupancy and hence the response regime to ionising radiation. In particular, although the total stopping power is size independent for same mass samples, the use of large NCs leads to the deposition of larger amounts of energy inside each particle, which consequently results in a higher *LY*; at the same time, large sizes reduce the AR efficiency, increasing the efficiency of the *XX* decay with beneficial effects on the timing performance. In fact, the generally high $\Phi_{AR}$ in CsPbBr$_3$ NCs leads to a scintillation process in which the *X* decay represents the major contribution, resulting in the *LY* being determined by the luminescence quantum yield, highlighting the crucial role of the surface chemistry for defect passivation and for ensuring the preservation of the optical properties of NCs after embedding in host matrices. On the other hand, the *XX* component is the key to accelerating $\tau_{EFF}$ and achieving high temporal resolutions for ToF applications, so optimising its yield, especially in larger particles, is paramount. More generally, the agreement found between the theoretical description and the experimental data is particularly relevant as it demonstrates the non-trivial possibility of predicting the main features of a NC scintillator from first principles - all the parameters in the equations are experimentally measured and the model therefore anticipates the actual behaviour - thus providing an extremely powerful tool for directing research towards the optimisation of key material parameters. Furthermore, the theory-experiment agreement validates the approximations of neglecting trionic contributions to scintillation and considering the totality of high-order excitons as biexcitons, further simplifying the prediction of actual performance. These results therefore fill a gap in the understanding of the scintillation process at the nanoscale and provide useful guidelines for specific particle engineering as the forthcoming generation of fast and efficient scintillators. Finally, we anticipate that the independence of the total stopping power from the particle size and number, but with the same total mass suggests that the differences in behaviour between isolated NCs with different dimensions may be reduced in composite nanoscintillators with a high density of NCs, where the electron shower released by one particle may act as a secondary excitation source of other NCs, giving rise to cascading scintillation phenomena. The study of this regime requires a specific treatment, which will be dealt with in a separate study.

**Acknowledgments**

This work was funded by Horizon Europe EIC Pathfinder program through project 101098649 – UNICORN, by the PRIN program of the Italian Ministry of University and Research (IRONSIDE project), by the European Union—NextGenerationEU through the Italian Ministry of University and Research under PNRR—M4C2-I1.3 Project PE_00000019 "HEAL ITALIA", by European Union's Horizon 2020 Research and Innovation programme under Grant Agreement No 101004761 (AIDAINNOVA). This research is funded and supervised by the Italian Space Agency (Agenzia Spaziale Italiana, ASI) in the framework of the Research Day "Giornate della Ricerca Spaziale" initiative through the contract ASI N. 2023-4-U.0

# Size-dependent multiexciton dynamics governs scintillation from perovskite quantum dots

Andrea Fratelli[1], Matteo L. Zaffalon[1], Emanuele Mazzola[2], Dmitry Dirin[3], Ihor Cherniukh[3], Clara Otero Martínez[4], Matteo Salomoni[2], Francesco Carulli[1], Francesco Meinardi[1], Luca Gironi[2,5], Liberato Manna[4*], Maksym V. Kovalenko[3*], Sergio Brovelli[1,5*]

[1] *Dipartimento di Scienza dei Materiali, Università degli Studi di Milano-Bicocca, Via R. Cozzi 55, 20125, Milano, Italy*

[2] *Dipartimento di Fisica, Università degli Studi di Milano-Bicocca, Piazza della Scienza 3, 20126, Milano, Italy*

[3] *Department of Chemistry and Applied Bioscience, ETH Zürich, Zürich, Switzerland.*
*Laboratory for Thin Films and Photovoltaics and Laboratory for Transport at Nanoscale Interfaces, Empa – Swiss Federal Laboratories for Materials Science and Technology, Dübendorf, Switzerland.*

[4] *Nanochemistry, Istituto Italiano di Tecnologia, Via Morego 30, Genova, Italy*

[5] *INFN - Sezione di Milano - Bicocca, Milano I-20126 – Italy*

## Methods

**Synthesis of $CsPbBr_3$ nanocrystal quantum dots.**

*Chemicals.* $Cs_2CO_3$ (99.9%, Sigma-Aldrich), diisooctylphosphinic acid (DOPA, 90%, Sigma-Aldrich), $PbBr_2$ (99.999%, Aldrich), trioctylphosphine oxide (TOPO, 90%, Strem), n-octane (97%, Acros Organics), hexane (96%, Scharlau), Oleic acid (OA, 90%, Sigma-Aldrich), lecithin (≥97%, from soy, Roth); mesitylene (99%, Thermo Scientific Chemicals); phosphorus(V) oxychloride (99%, Sigma-Aldrich); 2-aminoethan-1-ol (≥99.0%, Sigma-Aldrich); acetic acid (>99.8%, Sigma-Aldrich); triethylamine (99%, Sigma-Aldrich); 2-octyl-1-dodecanol (97%, Sigma-Aldrich).

*Stock solutions.* $PbBr_2$-TOPO stock solution was prepared by dissolving 367 mg $PbBr_2$ and 2.15 g TOPO in 5 ml n-octane at 120 °C, followed by cooling down and dilution with 20 ml hexane. Cs-DOPA stock solution was prepared by reacting 100 mg $Cs_2CO_3$ with 1 ml DOPA in 2 ml n-octane at 120 °C, followed by cooling down and dilution with 27 ml hexane. For mixed-halide NCs, $ZnCl_2$-TOPO and $ZnI_2$-TOPO stock solution for anion exchange were prepared by dissolving 136 mg $ZnCl_2$ or 319 mg $ZnI_2$ with 1.933 g TOPO in 2.5 ml octane at 120 °C, followed by cooling down and dilution with 7.5 ml hexane. 2-ammonioethyl 2-octyl-1-dodecyl phosphate ($C_8C_{12}$-PEA) ligand was synthesized as reported in Ref. *Nature* **626**, 542–548 (2024).

*Synthesis.* NCs were synthesized according to the modified $PbBr_2$-TOPO method described elsewhere (ref. *Science* **377**, 1406-1412 (2022)). For the synthesis of 6.8–9.8 nm, a 25 ml flask was loaded with hexane and $PbBr_2$-TOPO stock solution and stirred at 1200 rpm. Next, Cs-DOPA stock solution was injected, and the solution was stirred for a few minutes before the ligand was added ($C_8C_{12}$-PEA, 0.1M in mesitylene). The crude solution was rotary evaporated at room temperature until 5–7 ml left, washed with acetone (0.4–0.7 eq.), centrifuged at 12100 rpm, and redispersed in n-octane. To obtain mixed-halide NCs, $ZnCl_2$-TOPO or $ZnI_2$-TOPO stock solutions were added (equimolar to the $PbBr_2$) before the washing, followed by stirring for a few minutes. The synthesis details are given in the table below.

| Hexane | PbBr$_2$ | Cs-DOPA | Reaction time | C$_8$C$_{12}$-PEA 0.1 M | NC size |
|---|---|---|---|---|---|
| 8 ml | 2 ml | 1 ml | 3 min | 0.26 ml | 6.8 nm |
| 3 ml | 2 ml | 1 ml | 3 min | 0.26 ml | 8.6 nm |
| 3 ml | 3 ml | 1.5 ml | 2 min | 0.315 ml | 9.8 nm |

10.7–22.3 nm NCs were synthesized by a modified PbBr$_2$-TOPO approach (ref. *Science* 377, 1406-1412 (2022)) with a slow injection of Cs-OA and PbBr$_2$-TOPO precursors at a higher temperature. 10.7–13.5 nm NCs were capped with C$_8$C$_{12}$-PEA, washed with acetone, centrifuged at 10000 rpm, and redispersed in n-octane. 15.0-22.3 nm NCs were capped with a mixture of lecithin and C$_8$C$_{12}$-PEA (100:4), washed with ethyl acetate:acetone (1:2 by volume), centrifuged at 10000 rpm, and redispersed in n-octane.

**Transmission electron microscopy.** Transmission electron microscopy images were collected using a JEOL JEM2200FS microscope operating at 200-kV accelerating voltage.

**Monte Carlo simulation:** Monte Carlo simulations are fundamental to study the interaction of ionising radiation with our detectors, to optimise the experimental setup and to interpret the results of the measurements. Our Monte Carlo simulations are based on GEANT4, a toolkit developed at CERN for the simulation of the passage of particles through matter. Our simulations allow the accurate reproduction of the experimental apparatus consisting of ionising sources, scintillators, and light detectors, in addition to other components such as optical grease and Teflon. A configuration file allows for the easy modification of a wide range of parameters, for example the position, energy, and type of sources (alpha, gamma or beta), as well as the dimensions and type of scintillator. In particular, for these studies, cubic nanocrystals of CsPbBr$_3$ in a polystyrene case are simulated, with dimensions varying between 2 and 12 nm per side and initial electron energies spanning from 1 to 100 keV. NCs are randomly placed within the host and checks are performed to avoid overlaps. As the number of NCs increases, the computational time increases greatly and for this reason simulations with a high number of NCs are performed with parallel jobs on high performance servers. These simulations also take care of the generation and propagation of the scintillation light up to the light sensors.

**Optical Measurements.** Optical absorption measurements were measured in octane with an Agilent Cary 60 UV–Vis spectrophotometer. PL measurements were performed with a Varian Cary Eclipse fluorescence spectrometer, exciting the samples with a 3.5 eV (355 nm) laser, and collecting the emitted light with a phototube. PLQY for every sample was obtained by comparison with a standard with the same absorbance at the excitation energy. Time-resolved PL were carried out using 3.06 eV (405 nm) ps-pulsed diode lasers (Edinburgh EPL405, ~70 ps pulses); the emitted light was collected with a phototube coupled to a Cornerstone 260 1/4 m VIS-NIR Monochromator (ORIEL) and a time-correlated single-photon counting unit (time resolution ~400 ps).

**Ultrafast transient absorption spectroscopy** measurements were performed on Ultrafast Systems' Helios TA spectrometer. The laser source was a 10 W Ytterbium amplified laser operated at 1.875 kHz producing ~260 fs pulses at 1030 nm and coupled with an independently tunable optical parametric amplifier from the same supplier that produced the excitation pulses at 3.1 eV (400 nm). After passing the pump beam through a synchronous chopper phase-locked to the pulse train (0.938 kHz, blocking every other pump pulse), the pump fluence on the sample was modulated using a variable ND filter. The probe beam was a white light supercontinuum.

**Radioluminescence measurements**. CsPbBr$_3$ NCs solutions were excited by unfiltered X-ray irradiation using a Philips PW2274 X-ray tube, with a tungsten target, equipped with a beryllium window and operated at 20 kV. At this operating voltage, a continuous X-ray spectrum is produced by a Bremsstrahlung mechanism superimposed to the L and M transition lines of tungsten, due to the impact of electrons generated through thermionic effect and accelerated onto a tungsten target. The RL was collected using a custom apparatus featuring a liquid-nitrogen-cooled, back-illuminated, and UV-enhanced charge-coupled device detector (Jobin Yvon Symphony II) coupled to a monochromator (Jobin Yvon Triax 180) with 600 lines mm$^{-1}$ grating.

**Time-resolved scintillation**: The time response of scintillation light was measured using a pulsed X-ray source composed by a 405 nm ps-pulsed laser (Edinburgh Instruments, EPL-405) hitting the photocathode of an X-ray tube from Hamamatsu (N5084) set at 40 kV. The emitted light was collected using a FLS980 spectrometer (Edinburgh Instruments) coupled to a PicoHarp 300 hybrid photomultiplier tube working in time-correlated single photon counting (TCSPC) mode (time resolution ~150 ps).

The RL dynamics were analysed in a least squares sense using a triple exponential function convolved with the instrument response function. The effective decay time ($\tau_{EFF}$) was calculated using the ratio of all the components according to

$$\tau_{EFF}^{-1} = \sum_{i=1}^{3} \frac{R_i}{\tau_i},$$

where $R_i$ is the relative weight computed as the integrated area of the specific decay.

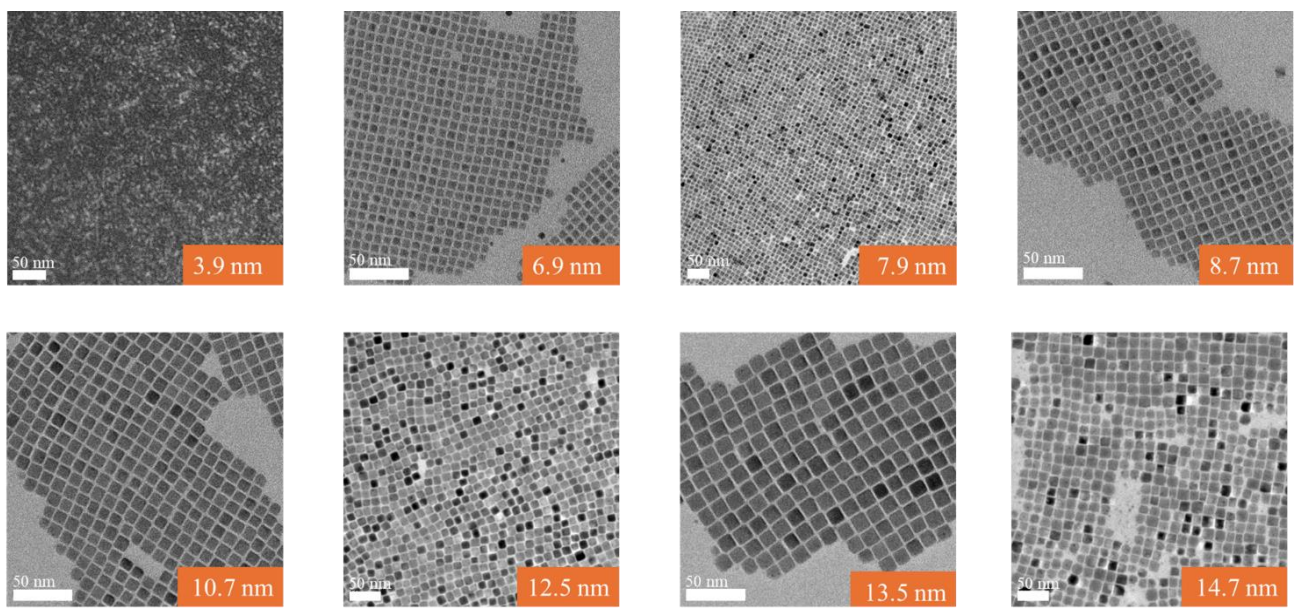

**Supplementary Figure 1**: representative HRTEM images for the set of NCs.

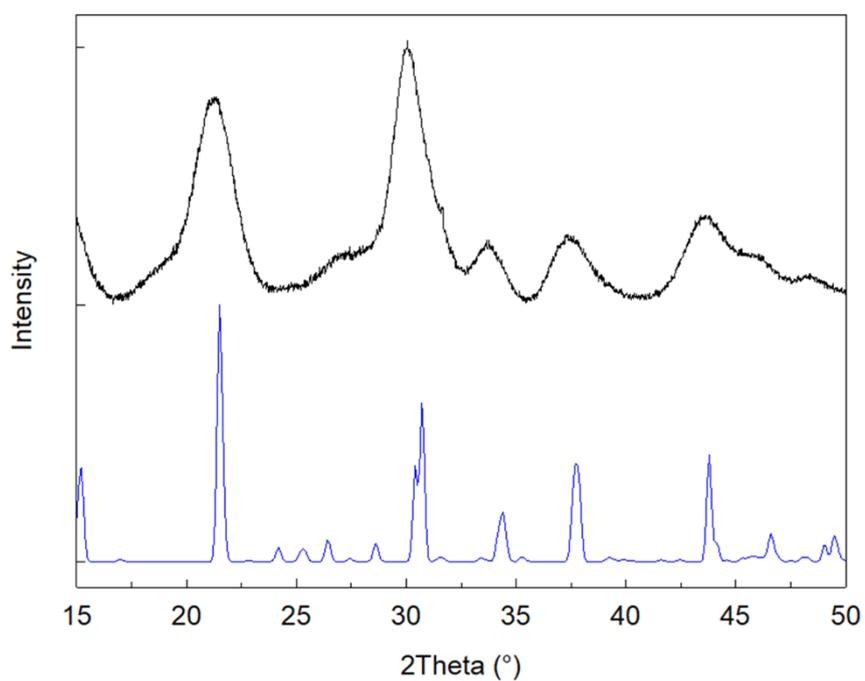

**Supplementary Figure 2**: X-ray diffraction pattern for the sample with d=3.5 nm NCs (black line), together with the calculated PXRD pattern for orthorhombic $CsPbBr_3$ (blue line, ICSD 97851).

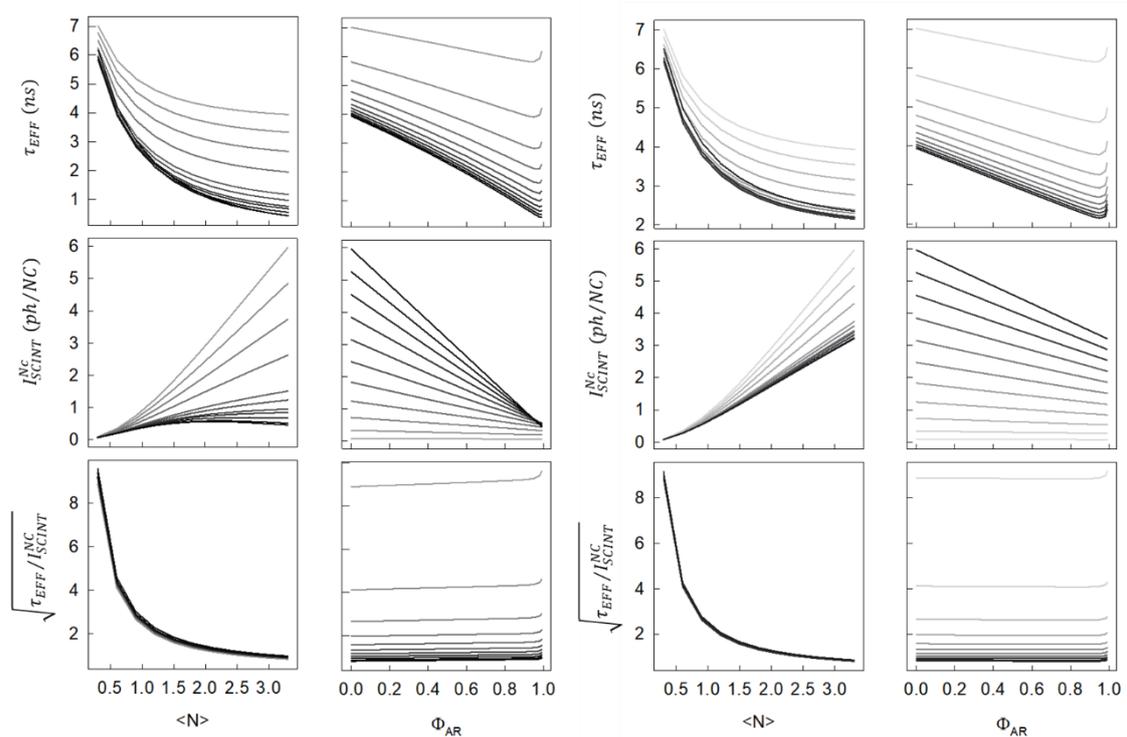

**Supplementary Figure 3**: simulation of the effective lifetime, scintillation intensity (per NC) and the square root of their ratio for $\xi = 0$ (ionizing AR) and $\xi = 1$ (non-ionizing AR).

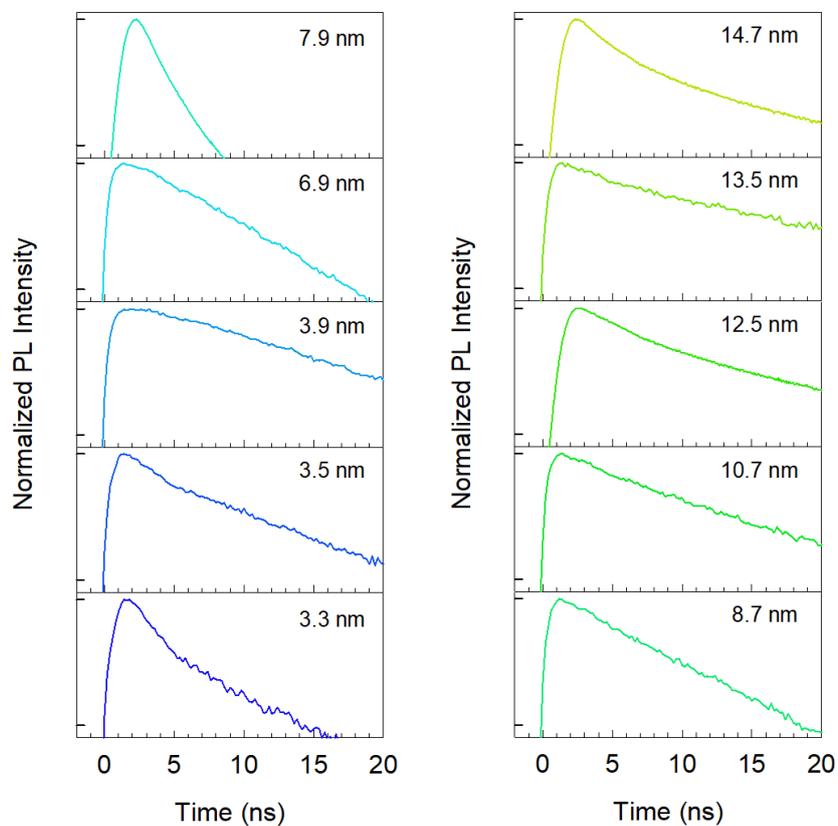

**Supplementary Figure 4**: time-resolved photoluminescence decays.

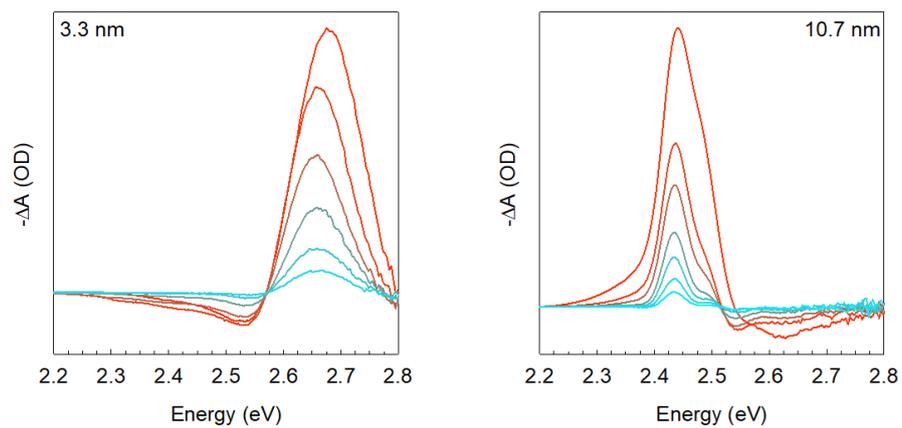

**Supplementary Figure 5**: transient absorption bleach spectra for the 3.3 nm and 10.7 nm NCs.

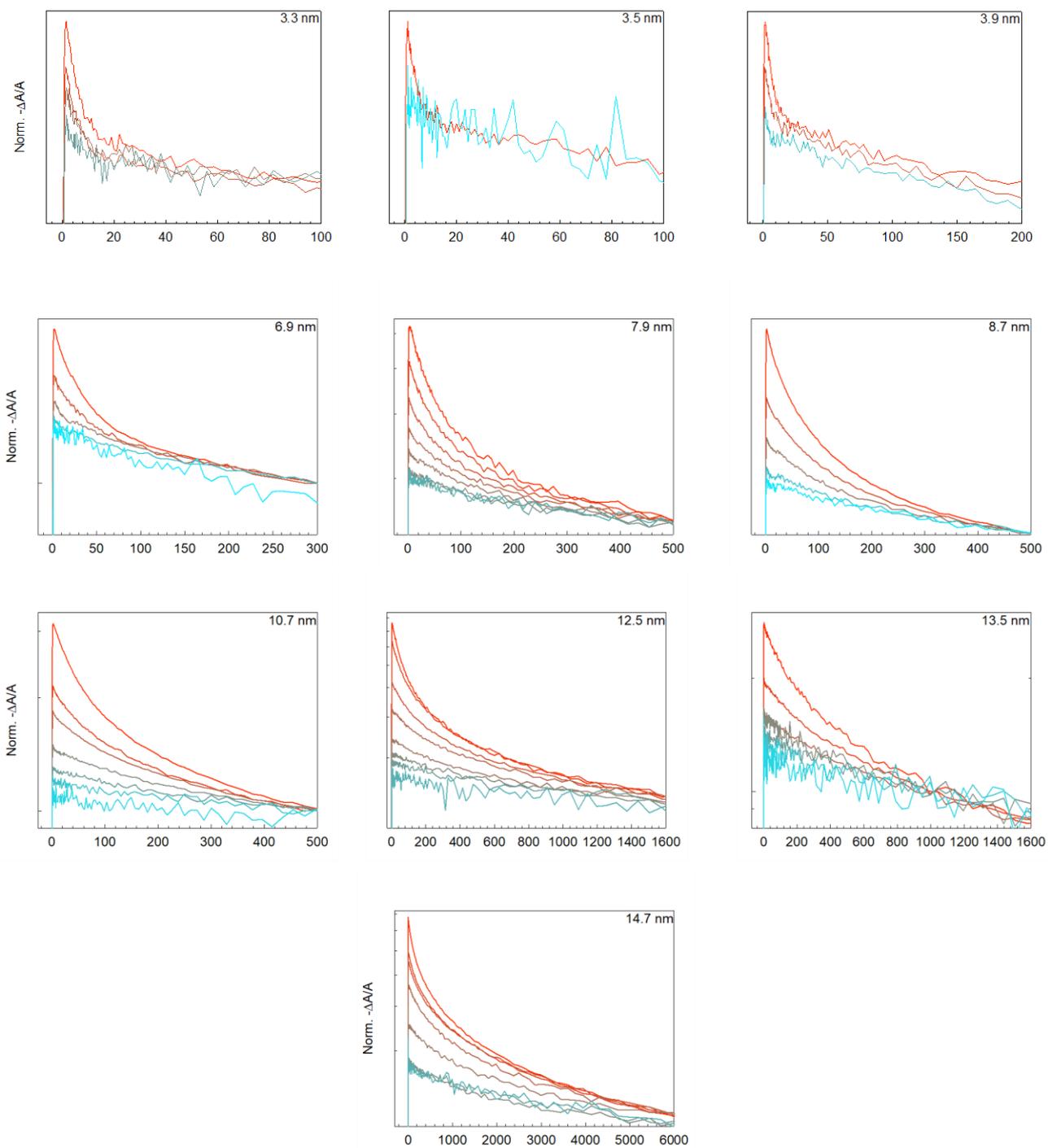

**Supplementary Figure 6**: transient absorption dynamics. The blue/red lines correspond to low/high fluences.

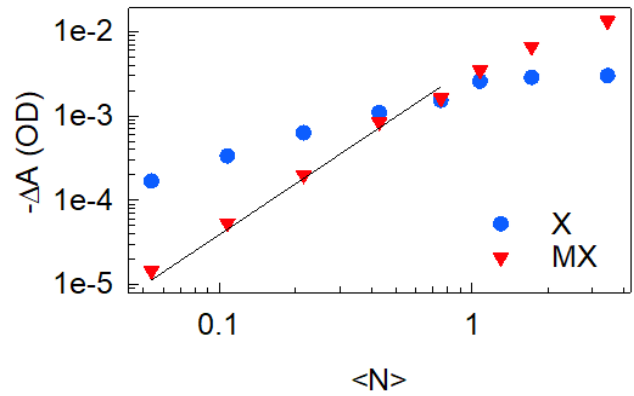

**Supplementary Figure 7**: Single exciton and multiexciton intensity with respect to the mean number of excitons per NC, for the sample with $d=13.5$ nm. In the low excitation regime ($\langle N \rangle < 1$), the multiexciton intensity results $\sim \langle N \rangle^2$.

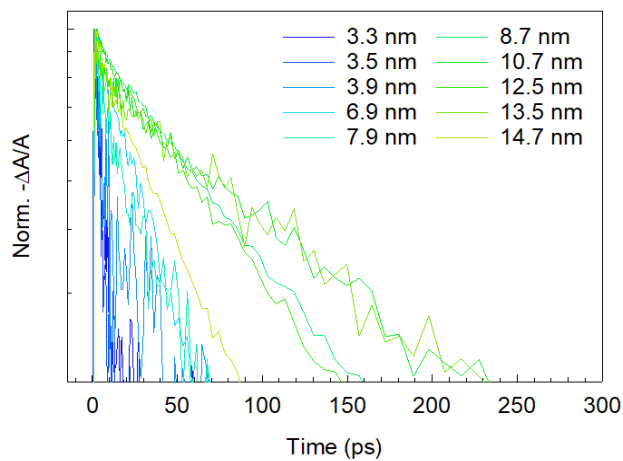

**Supplementary Figure 8**: Biexciton component, obtained by progressive subtractions, for the entire set of NCs, exhibiting a lifetime linearly increasing with respect to the NC volume.

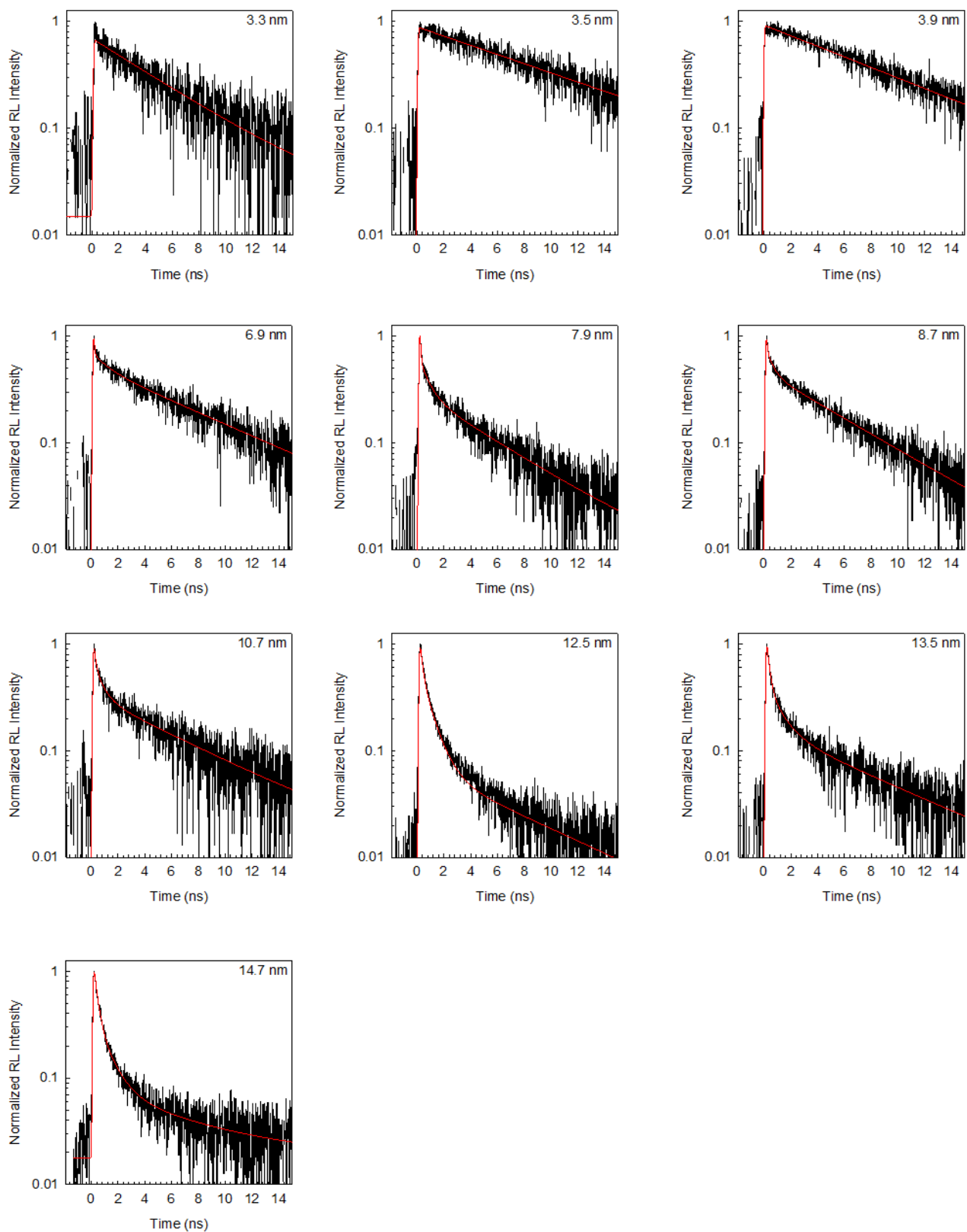

**Supplementary Figure 9**: Time-resolved scintillation decays (black lines) along with their fit (red lines).